\documentclass{appolb}
 \usepackage[dvips]{graphicx}
 \usepackage{amsmath}
 \usepackage{amsfonts}
 \usepackage{amssymb}

 \usepackage{color}

\newcommand{\df}{\ {\overset {\rm def} =}\ }

\newcommand{\dril}[2]{{{\rm d} {#1}} / {{\rm d} {#2}}}

\allowdisplaybreaks[4]

\begin{document}

\eqsec  

\title{Cosmological models and misunderstandings about them
}

\author{Andrzej Krasi\'nski
\address{N. Copernicus Astronomical Center, Polish Academy of Sciences \\
Warsaw, Poland}
}

\maketitle

\begin{abstract}
Advantages of inhomogeneous cosmological models that are exact solutions of
Einstein's equations over linearised perturbations of homogeneous models are
presented. Examples of effects that can be described in the inhomogeneous ones
are given: the non-repeatable light paths, the observed anisotropies in the
cosmic microwave background, the redshift drift and the maximum diameter
distance. Criticisms of inhomogeneous models that are based on misunderstandings
or fallacious reasonings are pointed out and corrected; these include the ``weak
singularity'', the positivity of deceleration ``theorem'', the ``pathology'' of
redshift behaviour at the ``critical point'' and the alleged necessity of the
bang time to be constant.
\end{abstract}

\PACS{PACS numbers come here}

\section{What is ``cosmological principle''?}

\setcounter{equation}{0}

The ``cosmological principle'' derives from Copernicus' discovery that can be
stated as follows: when the origin of coordinates is placed in the centre od the
Sun, the description of the motions of planets becomes simpler. In later
centuries further discoveries indicated that the position of the Sun in the
Universe is not in any way privileged. Ultimately, this conclusion assumed a
fundamentalist form: all positions in space are equivalent; every observer will
see the same large-scale image of the Universe.

This ``cosmological principle'' is not a summary of knowledge based on
observations, but a postulate. Just as all other hypotheses, it requires
observational verification.

Progress in observing technology, with still farther regions coming into view,
produced no justification for this principle: only more structures were becoming
visible. Nevertheless, we are told that the Universe is homogeneous ``at a
sufficiently large scale''. The definition of this ``sufficient scale'' is far
from precise (``a few'' hundred megaparsecs). This is the size of the
``fundamental cell'' of the Universe, which should be repetitive -- but it is so
large that details of mass distribution at its edges and beyond are fuzzy.

\section{Why consider generalised cosmological models?}

\setcounter{equation}{0}

Traditionally, structures are described by solutions of Einstein's equations
linearised around homogeneous models. This method has problems:

1. It is impossible to determine the radius of convergence of a series of
approximations when we know only the first and sometimes the second term -- as
is always the case in cosmology.

2. In practice, one demands that the parameter of a perturbative calculation is
smaller than 1 (but this is not a sufficient condition). In cosmology, there are
two such parameters: the density contrast $\Delta \rho / \rho_b$ and the
curvature contrast $\Delta R / R_b$ ($\rho_b$ is the mass density in the
background model, $\Delta \rho$ is the difference between the value of density
at the location considered and $\rho_b$, $R_b$ is the curvature of a
3-dimensional space of constant time in the background model, $\Delta R$ is the
analogue of $\Delta \rho$). \textit{Both} must be small. The curvature contrast
alone is not any measure of goodness of approximation \cite{PlKr2006}. But
$\left|\Delta \rho / \rho_b\right| < 1$ is not fulfilled in most objects
considered in cosmology:
 \begin{center}
 \begin{tabular}{|c|c|c|c|c|c|c|}
 \hline \hline
 \  & star & globular & galaxy & Virgo & Great & void \\
 \ & \  & cluster & \  & cluster & Attractor & \ \\
 \hline
$\Delta \rho / \rho_b$ & $1.5 \times 10^{29}$ & $2 \times 10^5$ & $6 \times
10^4$ & $190$ & $0.6$ & $-0.9$ \\
\hline \hline
 \end{tabular}
 \end{center}

3. Among the solutions of the linearised Einstein equations there are such that
are not approximations to any exact solution \cite{BrGi1999}.

Inhomogeneous models are not alternatives, but {\em generalisations} that reduce
to the traditional ones in the limit of spatial homogeneity. One should not
expect that observations will tell us that one class is fine, and the other
should be rejected. The relation is similar to that between a globe and a map of
a region of the Earth. A globe portrays the Earth as a perfect sphere, but a map
of a small area will show mountains and other features.

\section{Geometry of the cosmological models}

\subsection{The Robertson -- Walker (R--W) models (Fig. 1)}

\setcounter{equation}{0}

The metric of this model follows from {\em assumed symmetries of spacetime}:
\begin{equation}\label{3.1}
{\rm d} s^2 = {\rm d} t^2 - S^2(t) \left[\frac {{\rm d} r^2} {1 - k r^2}\ + r^2
\left({\rm d} \vartheta^2 + \sin^2\vartheta {\rm d} \varphi^2\right)\right].
\end{equation}
If the matter in the model has zero pressure, then $S(t)$ obeys
\begin{equation}\label{3.2}
{S,_t}^2 = 2G M / (c^2 S) - k + \Lambda S^2 / 3,
\end{equation}
where $k$ and $M$ are arbitrary constants and $\Lambda$ is the cosmological
constant.

 \begin{figure}[h]
 \begin{center}
 ${}$ \\[15mm]
\includegraphics[scale = 0.4] {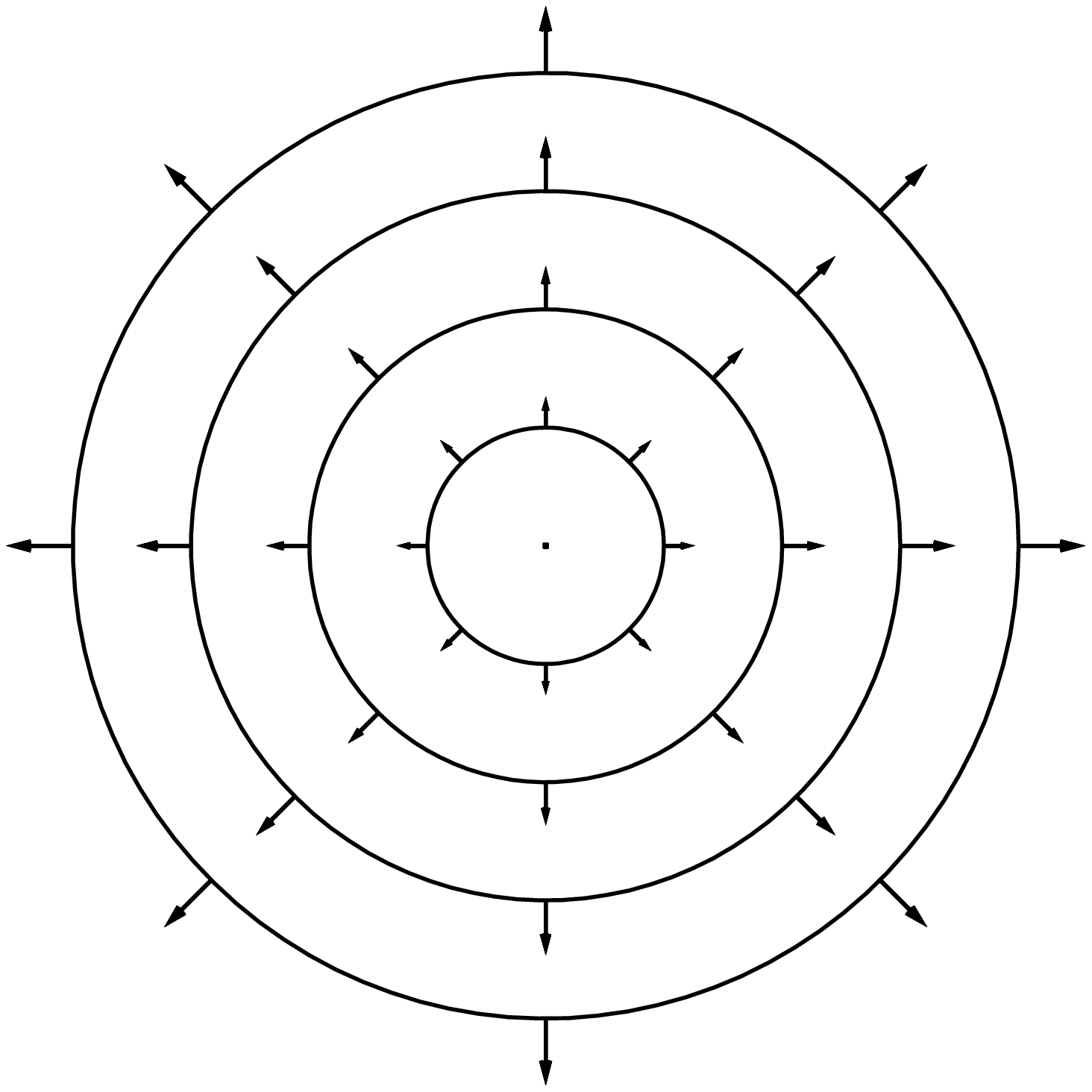}
 ${}$ \\[-10mm]
 \includegraphics[scale = 0.4]{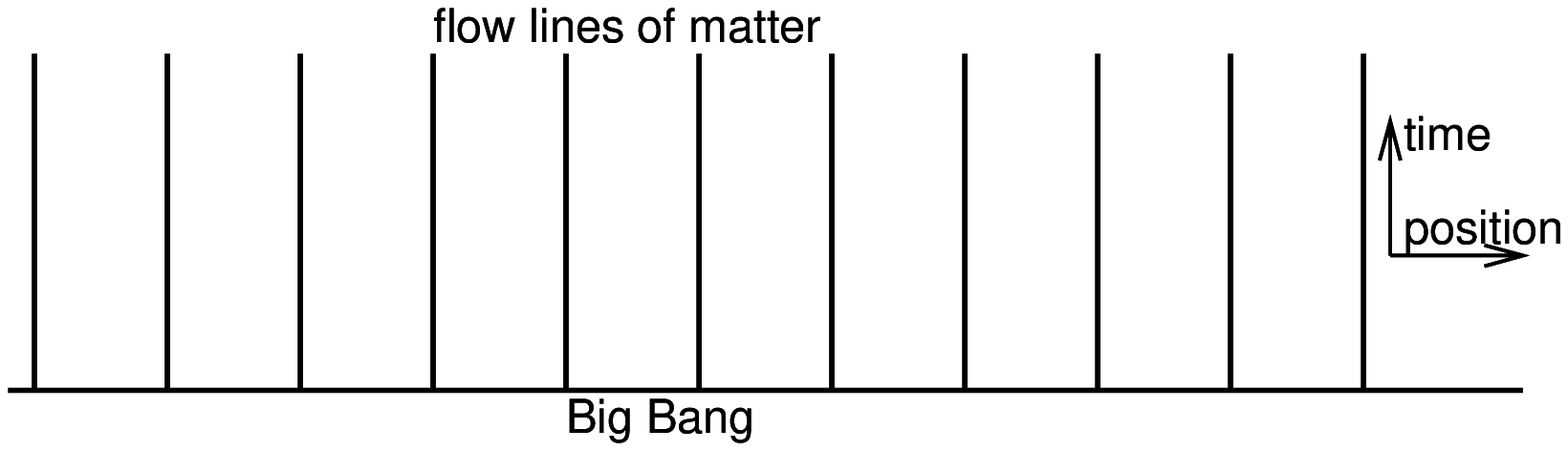}
 ${}$ \\[-5mm]
 \caption{
Expansion in the R--W models. {\bf Upper picture:} The velocity of expansion is
proportional to the distance from the observer at any fixed instant, but changes
with time. {\bf Lower picture:} The initial explosion occurs simultaneously in
the coordinates of (\ref{3.1}) $\Longrightarrow$ all matter particles have the
same age at any later instant. }
\end{center}
\end{figure}

The {\em redshift} $z$ is defined by
 \newpage
\begin{equation}\label{3.3}
z = \frac {\mbox {\rm emitted\ frequency}} {\mbox {\rm observed\ frequency}} - 1
\equiv \frac {\nu_e} {\nu_o} - 1.
\end{equation}
For the R--W models the redshift is:
\begin{equation}\label{3.4}
z = S(t_o)/S(t_e) - 1.
\end{equation}

The {\em luminosity distance} between an observer at $(t, r) = (t_o, 0)$ and the
source of light at $(t_e, r_e)$ is defined as the flat-space distance to a
source that would give the same observed flux of radiation, corrected for the
recession velocity of the source. In the R--W models this is
\begin{equation}\label{3.5}
D_L = r_eS(t_e) (1 + z)^2.
\end{equation}

The observable quantities are $z$, the Hubble coefficient at $t_o$:
\begin{equation}\label{3.6}
H_0 = \left.S,_t/S\right|_{t = t_o}
\end{equation}
and three dimensionless parameters
\begin{equation}\label{3.7}
\left(\Omega_m, \Omega_k, \Omega_{\Lambda}\right) \df \frac 1 {3{H_0}^2}
\left.\left(8\pi G \rho_0, - 3 k/{S_0}^2, \Lambda\right)\right|_{t = t_o}
\end{equation}
that obey $\Omega_m + \Omega_k + \Omega_{\Lambda} \equiv 1$. In these variables:
\begin{equation}\label{3.8}
D_L(z) = \frac {1 + z} {H_0 \sqrt{\Omega_k}} \sinh \left\{\int_0^z \frac
{\sqrt{\Omega_k} {\rm d} z'} {\sqrt{\Omega_m (1 + z')^3 + \Omega_k (1 + z')^2 +
\Omega_{\Lambda}}}\right\}.
\end{equation}
This formula applies also with $\Omega_k < 0$ ($\sinh ({\rm i} x) \equiv {\rm i}
\sin x$) and $\Omega_k \to 0$; the latter is the current favourite model of a
majority of cosmologists.

\subsection{The Lema\^{\i}tre -- Tolman (L--T) model (Fig. 2)}

 \begin{figure}[h]
 \begin{center}
${}$ \\[15mm]
  \includegraphics[scale = 0.5]{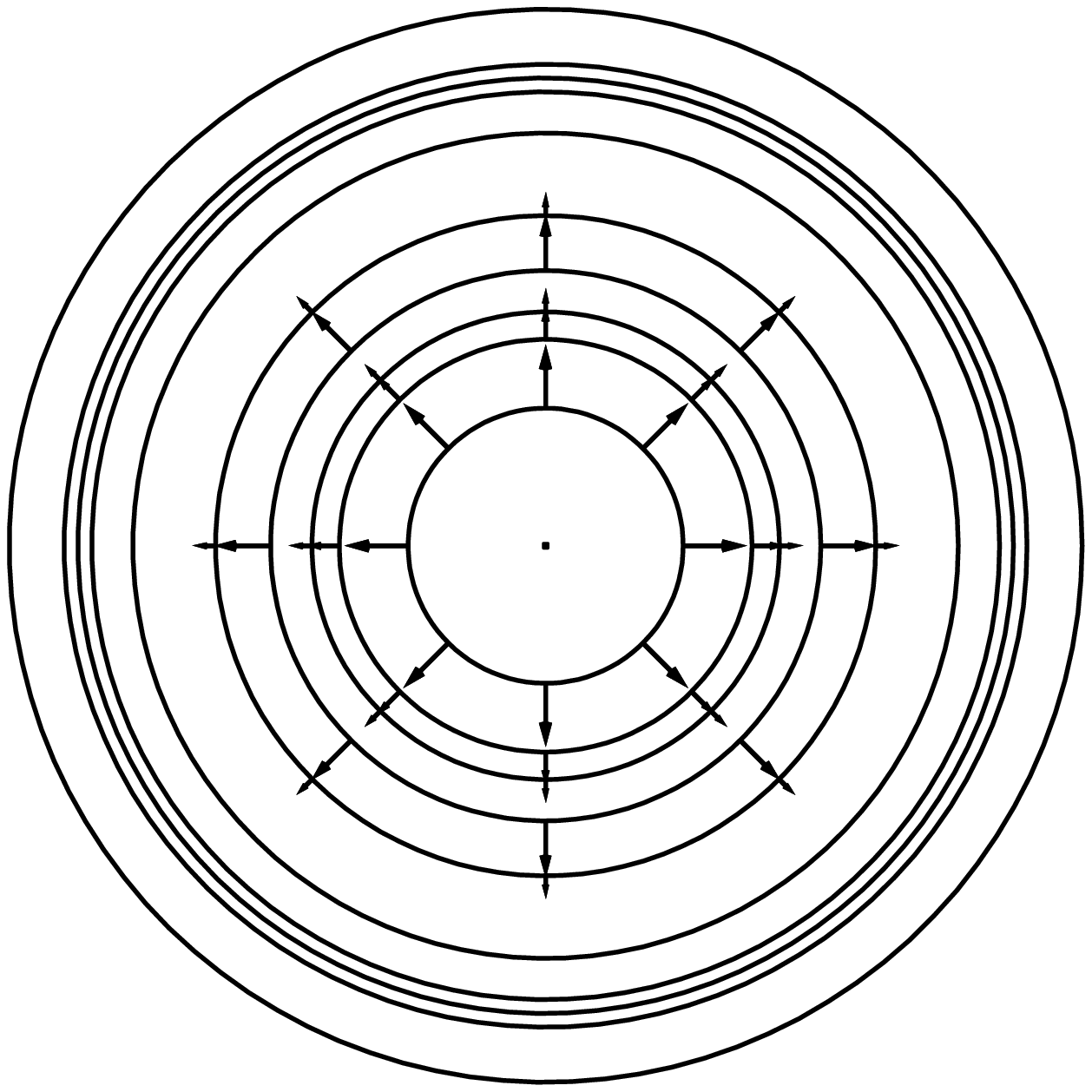}
${}$ \\[-10mm] \hspace{10mm}
 \includegraphics[scale = 0.4]{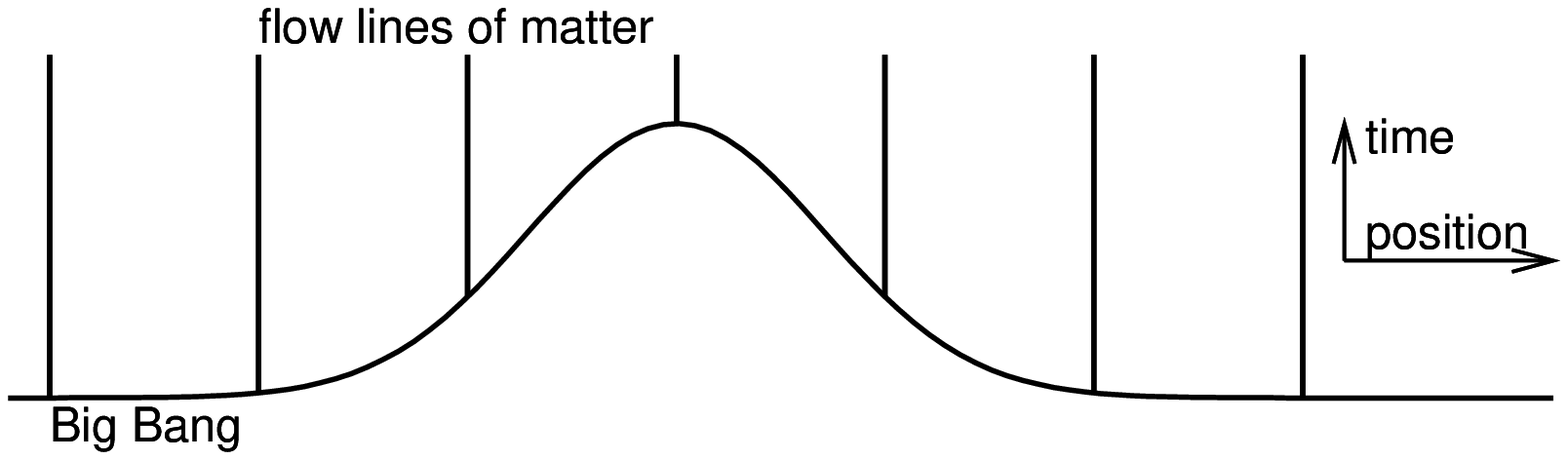}
 ${}$ \\[-5mm]
 \caption{
Expansion in the L--T model. {\bf Upper picture:} The velocity of expansion is
not correlated with the position of a matter shell. The spatial distribution of
velocity is an arbitrary function of $r$. {\bf Lower picture:} The initial
explosion is, in the coordinates of (\ref{3.9}), non-simultaneous
$\Longrightarrow$ the age of matter particles depends on $r$. The ``timetable''
of the initial explosion is a second arbitrary function of $r$. }
 \end{center}
\end{figure}

The metric of the Lema\^{\i}tre -- Tolman \cite{Lema1933, Tolm1934} model is
\begin{equation}\label{3.9}
{\rm d} s^2 = {\rm d} t^2 - \frac {{R,_r}^2} {1 + 2E(r)} {\rm d} r^2 - R^2(t, r)
\left({\rm d} \vartheta^2 + \sin^2 \vartheta {\rm d} \varphi^2\right),
\end{equation}
where $R(t, r)$ obeys (from the Einstein equations):
 \newpage
\begin{equation}\label{3.10}
{R,_t}^2 = 2 E(r) + 2M(r) / R - \Lambda R^2 / 3.
\end{equation}
The functions $M(r)$ and $E(r)$ are arbitrary. The integral of (\ref{3.10})
contains one more arbitrary function, $t_B(r)$ -- the ``timetable'' of the
initial explosion. For example, when $E = 0 = \Lambda$ the solution of
(\ref{3.10}) is
\begin{equation}\label{3.11}
R = (9M / 2)^{1/3} \left(t - t_B(r)\right)^{2/3}.
\end{equation}
The R--W limit of L--T is
\begin{equation}\label{3.12}
M= {\rm const} \cdot r^3, \qquad E = - kr^2/2, \qquad t_B = {\rm const}, \qquad
R = rS(t).
\end{equation}

The L--T model is spherically symmetric around one centre. It does not represent
the whole Universe, but a single structure embedded in an R--W background. One
R--W background can contain several L--T regions.

\section{Explaining away ``accelerated expansion'' of the Universe by
inhomogeneous matter distribution}

\setcounter{equation}{0}

The hypothesis of accelerated expansion of the Universe arose from observations
of type Ia supernovae. Such a supernova is the final stage of evolution of a
white dwarf in a binary system. The maximal absolute luminosity of all
supernovae of this class is assumed to be the same.

By measuring the redshift of these supernovae, one can calculate the distance to
them, {\em assuming that the Universe we live in is R--W with known parameters,
and so the Hubble law is exactly fulfilled} -- and then calculate the expected
flux of radiation through a unit surface area to be observed on the Earth. It
turned out that the actually observed maximal flux is smaller than expected, as
if the supernovae were farther from us than we thought. In order to explain this
discrepancy, {\em the previously used Universe model had to be modified}. {\em
Attempts to fit various R--W models to the observed luminosities} led to the
best fit achieved when $\Omega_k = 0$, $\Omega_m \approx 0.3$, $\Omega_{\Lambda}
\approx 0.7$.

A positive value of $\Omega_{\Lambda}$ means that the Universe has to expand
with acceleration. This gave rise to the puzzle of ``dark energy'' (that would
propel the acceleration) and to research programs aimed at solving it. But is
this the only possible explanation of the ``dimming of supernovae"?

The interpretation of observations is possible only when {\em the background
geometry of the space is pre-assumed}. The statements in italics in the text
above indicate the points at which this prior assumption
intervened.\footnote{This is in fact a vicious circle. We must assume a model to
interpret the observations, but then we use the observations to determine the
model. Efforts to break free from this circle are under way, but so far have not
led to generally usable results.}

This last remark must be exactly understood because its mistaken understanding
created a false legend. What we have to explain is the relation between the
observed luminosity of the type Ia supernovae and their redshifts, i.e. we have
to reproduce the function $D_L(z)$ in our model. {{\em The ``accelerated
expansion'' of the Universe is not an observed phenomenon, but an element of
interpretation of observations, forced upon us by the R--W models.}} If we can
re-create the observed $D_L(z)$ in a decelerating inhomogeneous model, then the
``accelerated expansion'' becomes an illusion.

This can be achieved using the L--T model \cite{INNa2002}. We assume that the
observer is at the symmetry centre and that $E / r^2 = E_0 =$ const, (the same
$E$ as in the R--W model). For $t_B(r)$ take the implicit definition:
\begin{equation}\label{4.1}
D_L(z) = \frac {1 + z} {H_0} \int_0^z \frac {{\rm d} z'} {\sqrt{\Omega_m (1 +
z')^3 + \Omega_{\Lambda}}},
\end{equation}
where $\Omega_m = 0.3$ and $\Omega_{\Lambda} = 0.7$, and $H_0$ is the present
value of the Hubble coefficient. The trick is that $z$ and $H_0$ are taken from
observations, but the definition of $D_L(z)$ is no longer (\ref{3.5}). Instead,
$D_L(z) = (1 + z)^2 R(t, z)$ is taken from an L--T model with $\Lambda = 0$, and
this defines $t_B(r)$. Equations for $t_B(r)$ can now be solved numerically.

Comparison with (\ref{3.8}) shows that (\ref{4.1}) defines the same relation
between $D_L$ and $z$ as in the ``standard'' R--W model with $\Omega_k = 0$,
$\Omega_m = 0.3$ and $\Omega_{\Lambda} = 0.7$. However, we achieved this with
$\Lambda = 0$, i. e. with decelerated expansion -- as dictated by the laws of
gravitation that we all know. Had we used the L--T model rather than R--W to
interpret the observations, there would be no need for the ``dark energy'' and
``accelerated expansion''.

\begin{figure}[h]
 \begin{center}
 \vspace{-10mm}
  \includegraphics[scale = 0.6]{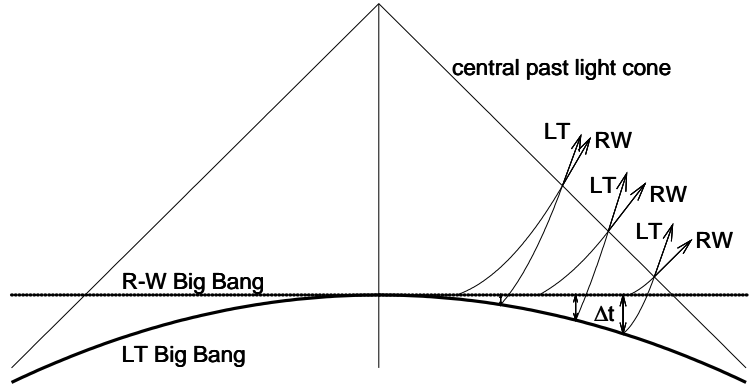}
 \vspace{-5mm}
 \caption{One of the ways of imitating accelerated expansion in the L--T model:
via a non-simultaneous Big Bang. Explanation in the text.
 } \label{imitaccel}
\end{center}
\end{figure}

Using this L--T model one can explain the ``dimming of supernovae'' in an
intuitively clear way. With a non-constant $t_B$ of suitable shape (see Fig.
\ref{imitaccel}), each L--T matter shell that intersects our past light cone is
older by $\Delta t$ than a $\Lambda = 0$ R--W shell that would intersect the
light cone at the same point. Therefore, at this intersection, the L--T shell
expands slower than an R--W shell would do. The $\Delta t$ increases toward the
past, and so does the difference between the expansion velocities. In this way
the L--T model imitates the acceleration of expansion relative to the $\Lambda =
0$ R--W model.

\section{Other results of relevance to inhomogeneous models}

 \begin{figure}[h]
 \begin{center}
  \includegraphics[scale = 0.7]{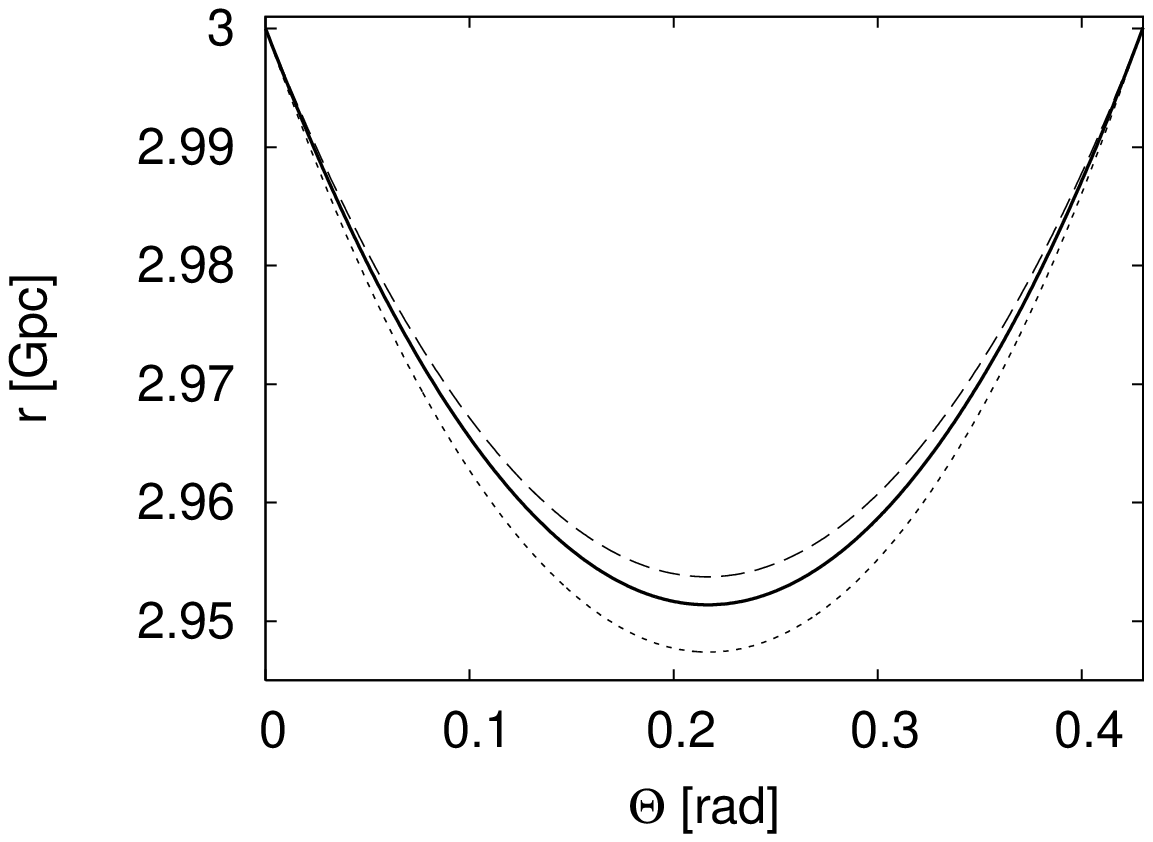}
\caption{Light rays sent between a given light source (at right) and a given
observer (at left) at different times do not proceed through the same sequence
of intermediate world lines. The graph shows three such rays projected on the
same space of constant time along the flow lines of the cosmic medium.
}\label{nonrep}
\end{center}
\end{figure}

\noindent $\bullet$ \hspace {3mm} {\bf Non-repeatable light paths}.

Generic light rays sent from the same source at different times to the same
observer pass through different sequences of intermediate matter particles
\cite{KrBo2010} (Fig. \ref{nonrep}). As a consequence, the observer should see
distant objects drift across the sky. Under most favourable conditions, the
drift rate would be 10$^{-7}$ to 10$^{-6}$ arc second per year, but should be
detectable after a few years of monitoring a given source, using devices that
are already under construction. This drift does not exist in the R--W models.

\noindent $\bullet$ \hspace {3mm} {\bf Anisotropies of temperature of the CMB
radiation.}

Inhomogeneities in mass distribution in the path of light rays cause directional
variations of temperature of the CMB radiation. Several investigations
\cite{BKHC2010} showed that no variations larger than $10^{-6} - 10^{-5}$ should
be expected. This agrees with the measured values. Thus, the high isotropy of
the CMB radiation does not imply that we live in an R--W model -- the
interaction between mass inhomogeneities and light is simply very weak.

\noindent $\bullet$ \hspace {3mm} {\bf Redshift drift}

As the Universe evolves, the redshifts of astronomical objects change with time.
For the $\Lambda$CDM model $\Delta z >0$ for $z<2$. For the giant void models
(see next section) $\Delta z < 0$ is expected for all $z$. Thus, a detection of
a negative redshift drift for all $z$ would be a proof against dark energy.
However, the converse is not true, as there are Gpc-scale inhomogeneous models
that also have $\Delta z >0$ for low $z$ \cite{YoKN2010}.

\noindent $\bullet$ \hspace {3mm} {\bf Maximum of the diameter distance}

In every model that begins with a Big Bang (R--W included!) the observed angular
diameter of distant objects decreases with distance up to a certain location,
and then increases as the observations approach the Big Bang. The position of
the minimum puts constraints on a model \cite{Hell2006} and is a consistency
check that may rule out some models.

\section{Errors and misconceptions}\label{errandmisc}

Astrophysicists are unusually tolerant toward a loose approach to mathematics.
Papers written in such a style planted errors in the literature, which were then
uncritically cited and came to be taken as established facts. In this section,
characteristic examples of misconceptions are presented (marked by $\huge
\blacksquare$) together with their explanations (marked by {\huge \bf $*$}).

\noindent ${\huge \blacksquare}$ \hspace {3mm} The L--T models that explain away
dark energy using matter inhomogeneities contain a {\em ``weak singularity''} at
the centre \cite{VFWa2006}, where the scalar curvature $R$ has the property
$g^{\mu \nu} R;_{\mu \nu} \to \infty$.

\noindent {\huge \bf $*$} \hspace {3mm} $g^{\mu \nu} R;_{\mu \nu} \to \infty$ is
not a singularity by any accepted criterion \cite{KHBC2010}. It only implies a
discontinuity in the gradient of mass density -- a thing common in Nature (e.g.
on the surface of the Earth). At the centre, $g^{\mu \nu} R;_{\mu \nu} \to
\infty$ implies a conical profile of density -- also a nonsingular
configuration.

\noindent ${\huge \blacksquare}$ \hspace {3mm} Decelerating inhomogeneous models
with $\Lambda = 0$ cannot be fitted to the $D_L(z)$ relation that implies
acceleration in $\Lambda$CDM. This is because the following equation prohibits
$q_4 < 0$ \cite{HiSe2005}
\begin{equation}\label{6.1}
H^2 q_4 = 4 \pi \rho / 3 + 14 \sigma^2 / 15
\end{equation}
($q_4$ is the deceleration parameter, $H$ is the Hubble parameter, $\rho$ and
$\sigma$ are the density and shear of the cosmic medium).

\noindent {\huge \bf $*$} \hspace {3mm} Equation (\ref{6.1}) is based on
approximations that are not explicitly spelled out \cite{KHBC2010}. An
approximate equation cannot determine the sign of anything. Refs.
\cite{BoHA2011,CBKr2010,INNa2002} provide explicit counterexamples to
(\ref{6.1}). If the approximations are taken as exact constraints imposed on the
L--T model, they imply the vacuum (Schwarzschild) limit. Moreover, the $q_4$ of
(\ref{6.1}), although it coincides with the deceleration parameter in the R--W
limit, is not a measure of deceleration in an inhomogeneous model.

\noindent ${\huge \blacksquare}$ \hspace {3mm} There is a {\em ``pathology''} in
the L--T models that causes the redshift-space mass density to become infinite
at a certain location (called {\em ``critical point''}) along the past light
cone of the central observer \cite{VFWa2006}.

\noindent {\huge \bf $*$} \hspace {3mm} The ``critical point'' is the apparent
horizon (AH), at which the past light cone of the central observer begins to
re-converge toward the past. This re-convergence had long been known in the R--W
models \cite{Elli1971,McCr1934}, and the infinity in density is a purely
numerical artefact -- a consequence of trying to integrate past AH an expression
that becomes 0/0 at the AH. Ways to handle this problem are known
\cite{LuHe2007,McHe2008,CBKr2010}.

\noindent ${\huge \blacksquare}$ \hspace {3mm} Fitting the L--T model to
cosmological observations, such as number counts or the Hubble function along
the past light cone, results in predicting a {\em huge void}, at least several
hundred Mpc in radius, around the centre (too many papers to be cited,
literature still growing).

\noindent {\huge \bf $*$} \hspace {3mm} The implied void is a consequence of
handpicked constraints imposed on the arbitrary functions of the L--T model, for
example a constant bang time $t_B$. With no a priori constraints, the giant void
is not implied \cite{CBKr2010}.

\noindent ${\huge \blacksquare}$ \hspace {3mm} The {\em bang time function must
be constant} because $\dril {t_B} r \neq 0$ generates decaying inhomogeneities,
which would have to be ``huge'' in the past, and this would contradict the
predictions of the inflationary models (private communication from the referees
of \cite{CBKr2010}).

\noindent {\huge \bf $*$} \hspace {3mm} The L--T models are not supposed to
apply prior to the emission of CMB because the pressure in them is zero. Thus,
their predictions for times close to the Big Bang cannot be taken literally.
Moreover, the occurrence of inflation is not in any way proven. Inflationary
models are just one of hypotheses that compete for observational confirmation.
Using them to justify or reject some other hypotheses is dogmatic and
un-scientific.

\section{A brief conclusion}

The theory of relativity has much more to offer to cosmology than the simplistic
R--W models found 90 years ago. Relativistic cosmology made a lot of progress
since then. The inhomogeneous models allow us to explain most of the observed
phenomena without introducing any ``new physics'' (like ``dark energy''). The
alleged deficiencies of the L--T models follow from hastily contrived reasonings
that contain errors in computation or in interpretation of the results.

{\bf Acknowledgements.} This article is based on several years of collaboration
with Krzysztof Bolejko, Marie-No\"{e}lle C\'el\'erier and Charles Hellaby. The
research for it was partly supported by the Polish Ministry of Education and
Science grant no N N202 104 838.

\end{document}